\documentclass[lettersize,journal]{IEEEtran}
\usepackage{amsmath,amsfonts}
\usepackage{algorithmic}
\usepackage{algorithm}
\usepackage{array}
\usepackage[caption=false,font=normalsize,labelfont=sf,textfont=sf]{subfig}
\usepackage{textcomp}
\usepackage{stfloats}
\usepackage{url}
\usepackage{verbatim}
\usepackage{subfig}
\usepackage{graphicx}
\usepackage{cite}
\usepackage[numbers,sort&compress]{natbib}
\usepackage{engord}
\usepackage{booktabs}
\usepackage[table]{xcolor}
\usepackage{colortbl}
\begin{document}

% \title{When and How Does Multimodality Benefit Robust Speech Recognition with Large Language Models?}
% \title{When and How Multimodal Integration Improves Language Model-based Speech Recognition}
\title{MLLM-based Speech Recognition:\\ When and How is Multimodality Beneficial?}

\author{Yiwen Guan, Viet Anh Trinh, Vivek Voleti, and Jacob Whitehill
        % <-this % stops a space
% \thanks{Manuscript received April 19, 2021; revised August 16, 2021.}% <-this % stops a space
\thanks{Yiwen Guan, Viet Anh Trinh, Vivek Voleti, and Jacob Whitehill are all with Worcester Polytechnic Institute, MA 01609, US (e-mail: yguan2@wpi.edu; vtrinh@wpi.edu; vsvoleti@wpi.edu; jrwhitehil@wpi.edu).}% <-this % stops a space
}


\markboth{IEEE TRANSACTIONS ON MULTIMEDIA,~Vol.~xxx, August~2021}%
{author name \MakeLowercase{\textit{et al.}}: Title}


\maketitle

\begin{abstract}

Recent advances in multi-modal large language models (MLLMs) have opened new possibilities for unified modeling of speech, text, images, and other modalities. Building on our prior work \cite{guan2024multi}, this paper examines the conditions and model architectures under which multiple input modalities can improve automatic speech recognition  (ASR) accuracy in noisy environments. Through  experiments on  synthetic and real-world data, we find that (1) harnessing more modalities usually improves ASR accuracy, as each modality provides complementary information, but the improvement depends on the amount of auditory noise. (2) Synchronized modalities (e.g., lip movements) are more useful at high noise levels whereas unsynchronized modalities (e.g., image context) are most helpful at moderate noise levels. (3) Higher-quality visual representations consistently improve ASR accuracy, highlighting the importance of developing more powerful visual encoders. (4) Mamba exhibits similar trends regarding the benefits of multimodality as do Transformers. (5) The input order of modalities as well as their weights in the loss function can significantly impact accuracy. These findings both offer practical insights and help to deepen our understanding of multi-modal speech recognition under challenging conditions.

\end{abstract}


\section{Introduction}
% \IEEEPARstart{I}{n} recent years, large language models (LLMs) have significantly advanced natural language processing tasks. Inspired by their remarkable reasoning and contextual understanding abilities, researchers have begun exploring their integration into multi-modal tasks, particularly in areas involving speech and vision. 
\IEEEPARstart{G}{iven} the success of large language models (LLMs) for natural language processing as enabled by their  reasoning and contextual understanding abilities, researchers are increasingly exploring how to develop multi-modal LLMs (MLLMs) to harness multiple input modalities, particularly in areas involving speech and vision \cite{comanici2025gemini,wu2024next,liao2025multimodal}. 
The evolution of LLMs -- specifically defined as large-scale autoregressive decoder-only models -- has led to the emergence of LLM-based automatic speech recognition (ASR) systems in the past few years \cite{trinh2024discrete,chang2024exploring,ma2025speech}. 
LLM-based ASR models typically adopt decoder-only architectures, taking either discrete units (e.g., extracted from HuBERT \cite{hsu2021hubert}) or continuous features (e.g., Log-Mel filterbank) as input, and producing text transcriptions as output.
While conventional speech processing methods consider only the audio itself, MLLM-based speech recognizers jointly model and process multiple input modalities, such as subject-matter context or visual cues like images and the speaker's lip movements \cite{wang2024comprehensive, chen2023x, yeo2025mms}. 
By exploiting complementary multi-modal information, these systems can sometimes reach higher accuracy in challenging conditions: noisy environments \cite{cappellazzo2025large, hong2023watch}, ambiguous words, long-tail vocabularies, and accented speech \cite{hu2024large, gabeur2022avatar, mu2024mmger}. For example, in noisy conditions, the scene context can resolve ambiguities between words like ``door" and ``dough", while lip movements help differentiate between ``night" and ``right"; or like the example in Fig. \ref{fig:datasets} that an image of a man in a fedora can help clarify the descriptive speech. 
They also benefit from LLM pre-training which endows them with linguistic information about which transcripts are more or less likely in a given context. 

MLLM-based approaches hold significant promise for next-generation ASR systems but they also bring new challenges:  While multiple input modalities may contain complementary information, they also increase the sequence length and complexity, making modeling more difficult. 
Each modality can be either beneficial or harmful in different situations that remain to be explored. 
Furthermore, it is still unclear what input formatting strategy -- including the order of inputs, positional encodings, and modality-specific loss weights --  is most effective in enabling the model to fully exploit multi-modal information. 
Moreover, Transformer-based models are inefficient when processing longer sequences due to their quadratic complexity attention mechanism, which sparks the research interest in faster architectures like Mamba \cite{gumamba}. 
% Due to the opacity of the internal mechanisms of the model, interpreting and understanding the process of modality integration and decision-making remains a challenge \cite{wang2024comprehensive}, so it becomes important to analyze the contribution of each modality. 
These challenges also raise questions about where researchers should focus to achieve better ASR performance: improving LLM architectures that can handle longer and more complex sequences, or enhancing modality-specific encoders to provide higher-fidelity representations?

In this paper, we investigate when and how multiple input modalities (speech audio, visual cues, and lip movements) are beneficial in multi-modal ASR. 
We systematically examine the conditions and model architectures under which each modality can improve ASR accuracy in noisy environments.
This article is an extended version of our earlier conference paper \cite{guan2024multi}, in which we initially analyzed the modality impact based on a multi-modal language model named DMLM \cite{trinh2024discrete}. DMLM (described in Section \ref{subsec:model}) is a discrete-token based unified model that processes multiple input modalities and performs various tasks with decoder-only backbone, including multi-modal ASR.  While our previous work investigated when and how each modality helps, here we provide a more in-depth analysis, expanded experiments, and several new contributions, as outlined below:

\begin{enumerate}
    \item Using both tightly-controlled synthetic as well as real-world datasets we examine the \emph{conditional} accuracy benefit of multi-modal inputs to MLLM-based ASR systems as a function of the auditory noise level. 
    \item We assess how the quality and efficiency of visual encodings affects ASR accuracy in MLLM architectures.
    \item We compare Transformer to Mamba as the MLLM backbone to assess the speed/accuracy trade-off as well as the difficulty of training.
    \item We compare different input strategies -- such as interleaving and swapping the modalities --  to demonstrate their impact on MLLMs, thus suggesting new directions for optimizing multi-modal systems.
\end{enumerate}

\section{Related Work}
\subsection{Multi-modal large language models}

Recent advances in MLLMs have demonstrated impressive capabilities in integrating information from multiple input streams to understand and reason across multiple modalities. 
Among these works, models like LLaVA \cite{liu2024improved} and Qwen-Audio \cite{chu2023qwen} focus on multi-modal comprehension tasks, such as answering questions grounded in visual scenes or spoken content, while other approaches, including NExT-GPT \cite{wu2024next} and MM-Interleaved \cite{tian2024mm}, extend unimodal LLMs to handle multi-modal generation tasks, such as rendering images or videos from textual descriptions or synthesizing speech from prompts.

% Meanwhile, 
A critical aspect of LLM design is its computational backbone,  and in recent years researchers have  explored replacing the most common Transformer architectures with more efficient ones.
 One notable alternative is the linear state-space model (SSM), such as Mamba \cite{gumamba} and Mamba-2 \cite{dao2024transformers}, which offers better scalability and efficiency than Transformers by avoiding the expensive attention mechanism.  
 Cobra \cite{zhao2025cobra} addresses the efficiency bottleneck of Transformer-based MLLMs by adopting pre-trained Mamba models and performs $3\times\sim4\times$ faster than the most efficient SOTA Transformer methods. U-Mamba \cite{ma2024u} leverages the powerful capability of SSMs in handling long sequences and proposes a CNN-SSM architecture to extract local features and long-range dependencies in biomedical image segmentation, which outperforms SOTA models in all tasks.

Understanding how MLLMs exploit multi-modal information is also crucial, so a growing body of work focuses on how different modalities interact within the model and how they contribute to overall performance \cite{wang2024comprehensive, ma2025speech}. They highlight the importance of effective modality fusion, cross-modality alignment, and robustness to modality degradation. Some work examines the effect of different model sizes, visual encoders, and modality connectors in ablation study to compare their effectiveness \cite{qiao2024vl}.
Other research raises research questions on how to quantify modality utility and how to formalize why modalities are useful or harmful \cite{liang2024foundations}.
However, they do not conduct systematic experiments or quantitatively analyze the contribution of each modality and rarely discuss the importance of secondary modalities, especially in noise and across datasets. These areas are closely aligned with our goal of analyzing and quantifying the contribution of each modality in the context of speech recognition under noisy conditions.

We also note the rise of diffusion-based language models, such as Diffusion-LM and LLaDA \cite{li2022diffusion, nie2025large}. 
As our focus is on autoregressive models, we omit diffusion-based language models from our study.

\subsection{Speech \& audio-visual systems}

Visual speech recognition (VSR), often known as automatic lip-reading, aims to recognize the speech content with the speaker's lip movements. As a broader combination of ASR and VSR, audio-visual speech recognition (AVSR) integrates both auditory and visual data, such as lip movements or unconstrained images, to improve speech recognition performance. Previous work \cite{hong2023watch} has analyzed the robustness of ASR, VSR, and AVSR models under different input corruption types for each modality. However, they only consider synchronized lip motions as visual information and overlook unsynchronized information like static images. Our work takes into account unsynchronized visual inputs and different model architectures.

Recent research illustrates the effectiveness of LLMs for ASR, VSR, and AVSR \cite{zhang2023speechgpt, zhang2025mamba, yeo2024visual, chu2023qwen, rubenstein2021audiopalm, cappellazzo2025large}. 
The integration of LLMs enables stronger multi-modal perception and contextual reasoning, which is particularly useful in noisy or ambiguous conditions.
In ASR, LLMs empower the system with stronger linguistic modeling, contextual adaptation, and generalization capabilities.
In AVSR, initiatives like Llama-AVSR \cite{cappellazzo2025large} incorporate powerful LLMs to more effectively combine auditory and visual cues under noisy conditions. 

While conventional AVSR systems primarily rely on synchronized audio and lip motion streams, the adoption of LLMs opens up more flexible and generalizable architectures for multi-modal speech recognition.
For instance, AVFormer \cite{seo2023avformer} incorporates unconstrained visual frames, while \cite{kim2025gesture} utilizes gestures as the visual modality, translating visual information into LLM-compatible representations for joint modeling. 
Zero-AVSR \cite{yeo2025zero} leverages the pre-trained LLMs' Roman-grapheme modeling capabilities to achieve zero-shot language recognition ability. 
Other approaches, including Llama-MTSK \cite{cappellazzo2025adaptive} and MMS-LLaMA \cite{yeo2025mms}, focus on optimizing audio-visual representations and introducing token compression techniques to reduce computational overhead without sacrificing accuracy. 
Despite these advancements, a fundamental question remains: how do individual modalities contribute to overall performance? A deeper understanding of modality contributions is essential for designing efficient, robust, and adaptive multi-modal systems for speech recognition.

\section{Methodology}

This work aims to understand the conditions in which  multiple input modalities can improve  speech recognition accuracy in MLLMs. We consider scenarios in which multiple modalities supplementary to the speech are available, including both \emph{synchronized} modalities such as lip movements (i.e., there is a temporal correspondence between the speech wave and the lip image sequence) as well as \emph{unsynchronized} modalities such as a picture of what the person is talking about (e.g., a lecture slide or a document). Moreover, we explore how the amount of auditory and/or visual noise can impact  the benefit of multimodality. Using both highly-controllable synthetic as well as real-world data, we systematically compare models with different modality combinations under varying noise conditions in order to quantify the accuracy benefits.

\subsection{Evaluation Metrics}

Let $\mathcal{M}=\{M_1, ..., M_{|\mathcal{M}|} \}$ be the set of all modalities (e.g., image, speech, etc.);  $m$ refers to one specific modality in $\mathcal{M}$. In our experiments we use three different metrics (WER, RB, and PS), which we report as percentages.

\subsubsection{Word Error Rate}
As a base metric to convey how accurate an ASR model is, we post-process the output transcriptions using the Whisper English text normalizer \cite{radford2023robust} and then calculate the word error rate (WER):
\begin{equation}
\label{wer}
\textrm{WER}=\frac{n_S+n_D+n_I}{N}
=\frac{n_S+n_D+n_I}{n_S+n_D+n_C}
\end{equation}
where $n_S$, $n_D$, $n_I$, $n_C$, and $N$ represents the number of substitutions, deletions, insertions, correct words, and reference words, respectively. Lower WER is better.

\subsubsection{Relative Benefit  of WER}
To evaluate how much the WER decreases by  adding one (or more) additional modalities to the audio-to-text task, we calculate the relative benefit (RB):
\begin{equation}
\label{rb}
\textrm{RB}(\mathcal{S})=\frac{\textrm{WER}(\{A\})-\textrm{WER}(\{A\}\cup \mathcal{S})}{\textrm{WER}(\{A\})}
\end{equation}
where $A$ denotes audio and $\mathcal{S} \subseteq \mathcal{M} \setminus \{A\} $ stands for the set of additional modalities. RB can be used as an assessment of the extent to which the additional modalities help the existing model. Larger RB values imply greater benefit from adding the extra modalities. Negative values indicate that the extra modalities hurt rather than help.

\subsubsection{Perceptual Score of WER}
To evaluate the influence of each modality in a  model that harnesses all modalities, we  compute the perceptual score (PS) \cite{gat2021perceptual} for each modality $m$:
\begin{equation}
\label{ps}
\textrm{PS}(m)=\frac{\textrm{WER}(\mathcal{M} \backslash \{m\} \cup \{\widetilde{m}\})-\textrm{WER}(\mathcal{M})}{\textrm{WER}(\mathcal{M})}
\end{equation}
where $\widetilde{m}$ indicates that  modality $m$ of each example has been randomly swapped with another example's modality $m$.
Larger PS indicates that $m$ is more influential.

% \subsubsection{Shapley Value of Relative Benefit}
% As a natural phenomenon of modality fusion, interactions between modalities always occur and it is hard to decouple the contribution of modalities. Hence, we consider calculating an RB-based Shapley value to examine the marginal contribution of each modality. In our task, the RB-based Shapley value is written in this format:
% \begin{equation}
% \label{phi}
% \phi_m=\frac{1}{\mathcal{|M|}}\sum_{\mathcal{S}\subset \mathcal{M}\backslash \{m\}} \binom{\mathcal{|M|}-1}{|\mathcal{S}|}^{-1}(\textrm{RB}(\mathcal{S}\cup\{m\}) - \textrm{RB}(\mathcal{S}))
% \end{equation}
% where $\mathcal{S}$ is any combination of modalities. Such Shapley value provides a possible approach to attribute benefits to each modality in the model. With a higher Shapley value, the modality is considered has a higher marginal contribution.

The relative benefit (RB) and perceptual score (PS) are subtly different: RB  estimates the performance improvement of adding new modalities to a model with fewer modalities; this is useful when deciding whether to use an audio-only model versus a multi-modal model.
In contrast, PS estimates the influence of each modality in a fully multi-modal model.

\subsection{Model Architecture and Pipeline}
\label{subsec:model}

\begin{figure}[t]
    \centering
    \includegraphics[width=1.0\columnwidth]{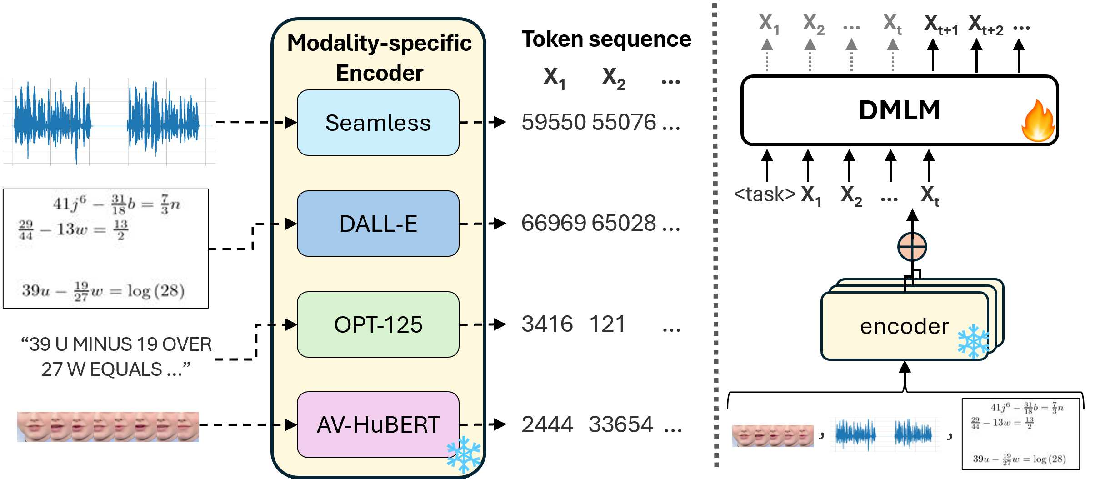}
    \vspace{-1.4em}
    \caption{ An overview of Discrete Multimodal Language Model (DMLM). The modality inputs are first tokenized, and then passed to DMLM for processing. The $\bigoplus$ in the figure represents modality vector concatenation.
    }
    \vspace{-1.0em}
    \label{fig:DMLM}
\end{figure}

\textbf{Model Architecture}: Our experiments are performed on a Discrete Multi-modal Language Model (DMLM, depicted in Fig. \ref{fig:DMLM}) \cite{trinh2024discrete}, a discrete token-based Transformer decoder-only model with an OPT-125M backbone \cite{zhang2022opt}
that was pre-trained on a variety of multimodal tasks including speech recognition, speech translation, image generation, and image captioning, using LibriSpeech-960, CVSS, CoVoST2, and COCO \cite{panayotov2015librispeech,jia2022cvss,wang2020covost,lin2014microsoft}. 
DMLM tokenizes the input data into discrete tokens using several frozen modality-specific encoders: we use Seamless \cite{barrault2023seamlessm4t} for audio, AV-HuBERT \cite{shilearning} for lip movements, and either DALL-E \cite{ramesh2021zero} or ViT 
\cite{dosovitskiy2020image} for images.
The input tokens of each modality are concatenated to form an input sequence, then the entire sequence is fed into OPT to be processed like text tokens. We use absolute positional encodings that are trained and added element-wise to the embedded input tokens. 
This pipeline is similar to other works like SpeechGPT \cite{zhang2023speechgpt} which takes multi-modal inputs and generates multi-modal outputs. As an alternative to OPT as the backbone, we also explored state space models using  Mamba-130M.

For multimodal models, an important consideration is the positional encoding that helps the model to find relevant information. Although we experimented with different positional embedding strategies (see Supplementary Materials Section II), we found that absolute encodings worked the best. 

%%%%%%%%%%%%%%%%% Figure %%%%%%%%%%%%%%%%%
%%%%%%%%%%%%%%%%% end figure  %%%%%%%%%%%%%

\textbf{Training Details}: Following prior work, we feed the model with training data in the format:
% ``[modality\_A\_modality\_B\_to\_text] \ldots (A tokens)\ldots \textless endofA\textgreater\ldots(B tokens)\ldots\textless endofB\textgreater\ldots(text labels)\ldots''. 
% ``(A tokens)\ldots \textless endofA\textgreater\ldots(B tokens)\ldots\textless endofB\textgreater\ldots(text labels)\ldots''. 
``\textless $m_1$ tokens\textgreater\textless $m_2$ tokens\textgreater\ldots\textless $m_n$ tokens\textgreater\textless text label (tokens)\textgreater''. 
The number of modalities ($n$) and the order of modalities vary in each task. We trained each model with AdamW \cite{loshchilovdecoupled} with $\beta$ =(0.9, 0.999), weight decay of 1e-4, learning rate of 1e-6 (unless specified). All experiments are conducted on a single NVIDIA A100 GPU with early stopping patience of 5.

\textbf{Inference Details}: The data are input in the format:  
% ``[modality\_A\_modality\_B\_to\_text] \ldots(A tokens)\ldots \textless endofA\textgreater\ldots(B tokens)\ldots\textless endofB\textgreater". 
% ``(A tokens)\ldots \textless endofA\textgreater\ldots(B tokens)\ldots\textless endofB\textgreater". 
``\textless $m_1$ tokens\textgreater\textless $m_2$ tokens\textgreater\ldots\textless $m_n$ tokens\textgreater". 
Then the model generates the corresponding text tokens autoregressively. To ensure determinism of the results, we set the temperature to 0 and use greedy decoding (no beam search) to generate tokens. Only the text part is used for  evaluation.

\subsection{Modality-Weighted Loss Function}
MLLMs trained autoregressively to predict the entire input sequence, which comprises multiple modalities, sometimes have hyperparameters  ($\lambda_S, \lambda_T, \lambda_I$, etc.) controlling how much each token type (speech, text, image, respectively) influences the loss function. Previous work \cite{trinh2024discrete} found a strong benefit, in terms of both training stability and testing accuracy, in  tuning these hyperparameters in a  multi-modal loss  of the form
\begin{equation}
\label{loss}
\mathcal{L} = \sum_{m\in \mathcal{M}} \frac{\lambda_m}{n_m} \mathcal{L}_{CE}(m)
\end{equation}
where $\mathcal{L}_{CE}(m)$ is the cross-entropy loss for predicting tokens and $n_m$ is the number of tokens in modality $m$. In particular, since $n_m$ can differ significantly over modalities, judicious tuning of the $\lambda$ can prevent any one modality from dominating the loss.
More subtly, previous work \cite{shilearning,afouras2018deep} has found that, in audio-visual speech recognition tasks, the audio signal can often dominate the model decisions because it is much easier for the model to associate the audio input with the lexical text output than lip reading. % (in our case, we also consider other visual tasks like image-to-text).
Hence, it is conceivable that a multi-modal ASR model may need a larger weight for non-speech modalities (image, lips)  than speech in order to ensure better learning of image-to-text token correspondences. %On the other hand, we also suspect that the model requires different configurations of modality loss weights during pre-training and fine-tuning.

In our experiments, we optimized the
%Based on previous work \cite{trinh2024discrete}, we pre-train the DMLM using their optimal modality loss weights ($\lambda_S$=0.25, $\lambda_T$=0.93, $\lambda_I$=0.25). We then explore the best combination of 
speech, text, and image loss weights  using a  grid search to minimize WER on a validation subset of the 3-Equations dataset (described below). This yielded an optimal hyperparameter configuration of $\lambda_S$=0.3, $\lambda_T$=0.5, and $\lambda_I$=0.5. %We keep this configuration in future fine-tuning tasks, and retain the original weights for pre-training.
Our experiments revealed that tuning modality loss weights was very influential in both Transformer-based and Mamba models. 
See Section \ref{subsec:modal_loss} and Supplementary Materials (Section III).

\section{Datasets}
Our experiments use one synthetic dataset (3-Equations) and 2 real-world datasets (SlideAVSR; Spoken Moments in Time).

%%%%%%%%%%%%%%%%% Figure %%%%%%%%%%%%%%%%%
\begin{figure*}[t]
    \centering
    \includegraphics[width=1.0\textwidth]{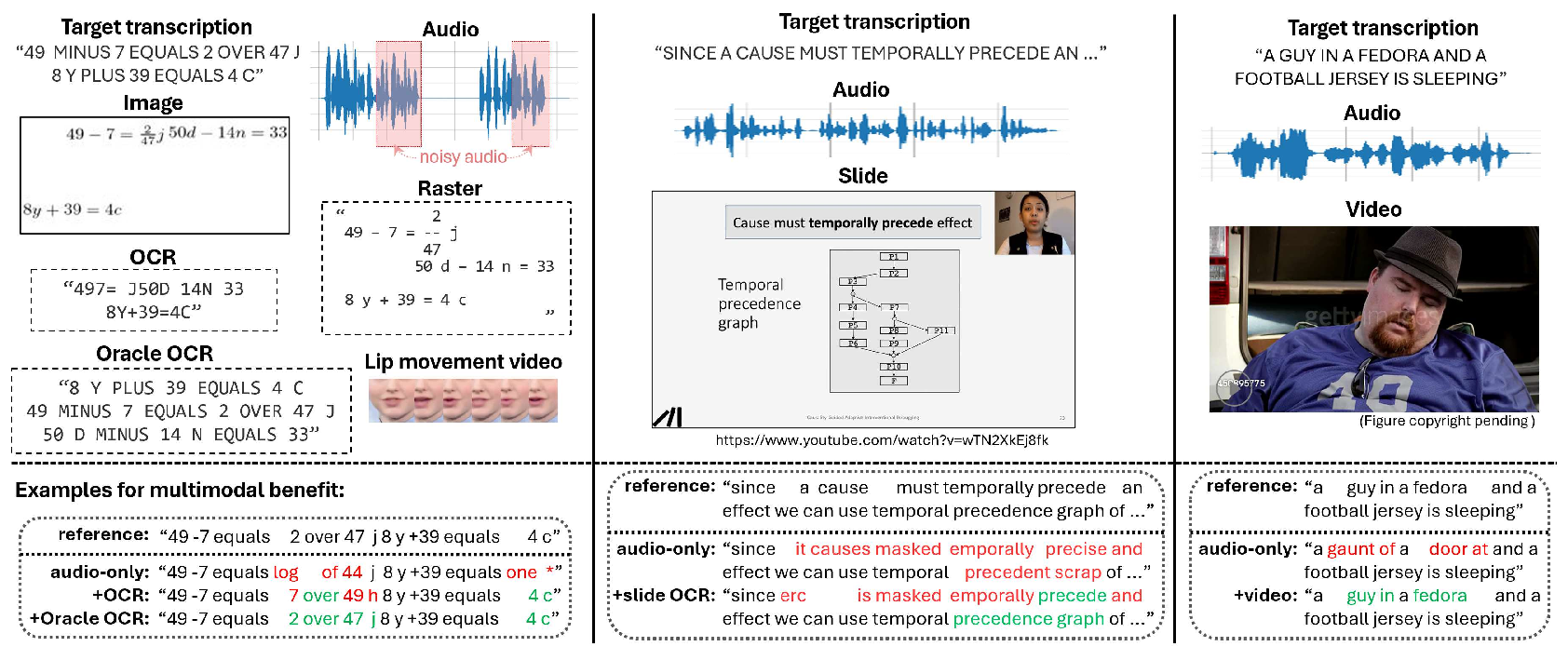}
    \vspace{-1.5em}
    \caption{ Datasets used in our work. From left to right: Examples from the 3-Equations \cite{guan2024multi}, SlideAVSR \cite{wang2024slideavsr}, and S-MiT \cite{monfort2021spoken}. For 3-Equations (left), we have different visual representations such as the image, OCR text, and raster representations. We provide one example for each dataset at the bottom that shows the help of incorporating multiple modalities in our experiments: ``reference" is the ground truth transcription, ``audio-only" is the output of an audio-only model (``audio$\rightarrow$text"), and ``+OCR" gives the output of the model trained on ``OCR+audio$\rightarrow$text" task.
    }
    \vspace{-1.0em}
    \label{fig:datasets}
\end{figure*}
%%%%%%%%%%%%%%%%% end figure  %%%%%%%%%%%%%

\subsection{3-Equations}
We created a synthetic multi-modal speech recognition dataset called 
3-Equations \cite{guan2024multi} to simulate scenarios in which multimodality is crucial to obtain  accurate speech transcription. It consists of speech utterances, synchronized lip movements, and image context. In each  utterance, a speaker reads 2 randomly selected mathematical equations that are displayed in an accompanying image containing 3 equations. We use the default male voice of pyttsx3 \cite{pyttsx3github} to synthesize the  voice.
To each utterance, MUSAN noise \cite{snyder2015musan} is added to the second half of each equation utterance so that only half of each utterance remains clean (see Fig. \ref{fig:datasets} left). %The clean first (clean) half of each spoken equation helps the model to find a correspondence between the audio and image modalities, while the second (noisy) half enables us to study how supplementary modalities improve ASR accuracy.
%We call this method of adding noise ``2-noise".
This noise setting encourages the model to learn from both visual and auditory modalities: without visual information (lip movements, images), the model will fail in noise; without the audio, the model may not have enough information about which 2 of the 3 equations were actually spoken  from the image. While our primary goal is  a tightly-controlled dataset in which we can manipulate noise, visual representation, and other factors, 3-Equations is also relevant to  educational applications such as multi-modal learning assistants.

In order to study the impact on ASR accuracy of the visual representation quality, we created a ``continuum'' of visual modalities: 
no visual input, image encoding, optical character recognition (OCR), 2-D raster, and oracle-OCR text.
For the image encoding, we used either DALL-E or ViT. For OCR, we applied EasyOCR \cite{easyocr} to each image. For oracle OCR, we concatenate the 3 equation sentences in each image, without filtering out the one that is not spoken. To similarly design a perfect 2-D representation that contains more spatial information than the OCR text, whilst still requiring conversion from math characters into English text, we created a 2-D text-based raster by placing characters from the equations onto a fixed-size text grid. With this new representation, the model is guaranteed to see the numbers and symbols in the equations, and it can potentially learn spatial information like image encoders. 
Hence, the continuum above can be considered as in an ascending order of visual representation quality. 
Fig. \ref{fig:datasets} (left) shows all visual modalities in 3-Equations. 
% Hence, the following sequence can be considered as a continuum in an ascending order of visual representation quality: 
% no image, image encoding, OCR, oracle OCR.

Altogether, the dataset consists of 10,000 examples, 8,000 for training and 1,000 each for validation and testing. In total it contains 25.2 hours of synthesized speech, with 20.2 hours for training and 2.5 hours each for validation and testing.

% %%%%%%%%%%%%%%%%% Figure %%%%%%%%%%%%%%%%%
% \begin{figure}[t]
%     \centering
%     \includegraphics[width=1.0\columnwidth]{figures/Images.png}
%     \vspace{-1.5em}
%     \caption{Illustration of different visual representations in the 3-Equations. From left to right, top to bottom: original image; DALL-E encoded 32$\times$12 tokens; OCR text; oracle OCR text; raster representation with 24$\times$36 characters (the number of characters is reduced in this figure for ease of display).
%     }
%     \vspace{-1em}
%     \label{fig:visual_encode}
% \end{figure}
% %%%%%%%%%%%%%%%%% end figure  %%%%%%%%%%%%%

\subsection{SlideAVSR}
SlideAVSR \cite{wang2024slideavsr} is an audio-visual dataset of videos collected from YouTube in which the speakers explain scientific papers; the dataset provides manually transcribed speech, synchronized slides, and preprocessed OCR keywords (see Fig. \ref{fig:datasets}, middle). The dataset contains many technical terms and the speakers have varying accents and speech intelligibility, making it difficult to transcribe accurately without the visual information (e.g., slides). It thus is a good choice to study how multi-modal architectures can handle auditory noise in real-world scenarios. 

Each utterance in SlideAVSR corresponds to one slide image with its processed OCR text. Some of the videos have talking faces, but since the coverage of talking face in this dataset is less than 50\%, we omit the lip modality in our experiments. 
We also add MUSAN noise with different signal-to-noise ratios (SNRs) to the entire audio to simulate broader range of noise conditions. This dataset comprises 245 hours of audio data, with 195 hours for training, 20 hours for validation and 30 hours for testing.

\subsection{Spoken Moments in Time (S-MiT)}
Spoken Moments in Time (S-MiT) \cite{monfort2021spoken} is a large-scale video caption dataset built from the Multi-Moments in Time \cite{monfort2021multi}; it consists of 500k short videos depicting a broad range of different events, such as chef cooking or bird singing, along with audio recordings describing them (see Fig. \ref{fig:datasets}, right). This dataset has a unique vocabulary size of over 50k, with high diversity and broad coverage examples.

While MLLMs with limited capacity may solely use static visual information gleaned from a single snapshot, stronger models can leverage more spatio-temporal information like motion. Using  S-MiT, we can investigate whether models can gain the ability to harness such dynamics. Hence, 
we  extracted 5 frames from each video uniformly over its duration. By comparing a visual modality consisting of all 5 frames to one containing only  the first frame of each video,
we can assess how much the motion is used beyond the static appearance.

Due to the input length limitation of the model we use in experiments, we subselect  S-MiT to include  examples whose audio duration is less than 5 seconds and video length less than 3 seconds, resulting in 56,385 examples for training, 692 for validation, and 146 for testing.

\section{Experiments: Noise \& Number of Modalities}
Multimodality can be a double-edged sword: With additional modalities, complementary information is provided, potentially leading to better accuracy. However, using more modalities increases the input length, which might increase the difficulty of finding relevant information and  actually degrade performance. Moreover, even when multimodal inputs do increase ASR accuracy, their benefit might derive not from complementary information but from implicit regularization: training an MLLM to predict the next token for every modality prevents the model from dedicating all its capacity to minimizing the training loss on the ASR text tokens.

In our experiments, we investigated whether using more modalities generally leads to higher ASR accuracy, as well as the conditions under which each modality is most effective. We explore conditions including acoustic noise, visual noise, and sequence noise. %We validate our findings on 3-Equations, SlideAVSR and S-MiT.
We trained models with all combinations of modalities (e.g., $I+A\rightarrow T$, $L+A\rightarrow T$, $I+L+A\rightarrow T$, etc.), then calculated the WER and RB of each model at different noise levels. Each model is fine-tuned on the 3-Equations training set containing a random mix of added MUSAN noise with an SNR in [20, 10, 5, 0, -5, -10, -20]. Then, each model was evaluated separately on an SNR-specific test set in [20, 10, 5, 2.5, 0, -5, -10, -20]. % -- this allows us to understand the impact of noise.

\begin{figure}[t]
    \centering
    \includegraphics[width=1.0\columnwidth]{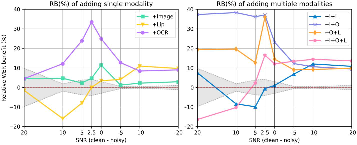}
    \vspace{-1.8em}
    \caption{RB (\%) of adding image, OCR, or lip modalities on 3-Equations 2-noise test set. The gray shading part represents the statistical error range (between RB+ and RB-) from our shuffled experiments. Left: adding single modality; Right: adding multiple modalities.
    }
    \vspace{-1.2em}
    \label{fig:3eq_rb_combine}
\end{figure}

To ensure that any performance gap between models is statistically reliable, we estimate for each SNR value the variability in WER of the \emph{same} model under a tiny perturbation (specifically, the order of examples across minibatches). This allows us to distinguish between meaningful WER differences versus just statistical anomalies. Specifically, the gray area in Fig. \ref{fig:3eq_rb_combine} between two dashed lines is the non-significant window due to just a mini-batch reshuffle. See Supplementary Materials (Section I) for more details.
% real and not just due to statistical noise, we estimate the variability 
% o ensure that the performance gap between experiments is caused by the addition of modality and not statistical noise. 
% Therefore, we conduct two experiments two times with different random shuffles of the mini-batches. We define $RB^+=\max(RB_1,RB_2)$ at each SNR as the upper bound of statistical error, and $RB^-=(-1)\times RB^+$ as the lower bound to evaluate significance. The gray area in Fig. \ref{fig:3eq_rb_combine} between two dashed lines is the insignificant window. Since most of our results are located outside the insignificant window, our experimental results and conclusions are not trivial.
% More details are included in supplemental materials.

Results on 3-Equations are shown in Table \ref{tab:3-eq-res-opt} and Fig. \ref{fig:3eq_rb_combine}.
As shown in the figure, most of the points on the curves are outside the non-significant window, indicating our results are non-trivial.  We also calculated WERs for each non-audio modality: For  $I\rightarrow T$, $L\rightarrow T$, $O\rightarrow T$, and $O_{oracle}\rightarrow T$ they were 79.9, 41.4, 76.5, and 60.5, respectively, which illustrate how the audio signal itself is crucial to obtaining low WER. We interpret the results in the subsections below.

%%%%%%%%%%%%%%% TABLE %%%%%%%%%%%%%%%%
\definecolor{lightgray}{gray}{0.9}
\begin{table*}[hbt]
    \caption{WERs and relative benefits of adding modalities to the Transformer-based model on 3-Equations, at different noise levels based on SNR. A, I, O, O$_{oracle}$, L, R, and T stand for audio, image, OCR, oracle OCR, lip, raster, and text, respectively.}
    \centering
    \setlength{\tabcolsep}{5pt}
    \resizebox{\textwidth}{!}{%
    \begin{tabular}{rccccccccc}
        \noalign{\hrule height 1.2pt}
        \multicolumn{1}{r|}{\textbf{Input Modalities}} & 20 & 10 & 5 & 2.5 & 0 & -5 & -10 & -20 & \multicolumn{1}{|c}{\textbf{Average}} \\
        \multicolumn{10}{c}{\cellcolor[HTML]{DEDCDC}WER $\downarrow$ (RB $\uparrow$)} \\
        \hline
        \multicolumn{1}{r|}{$A$} & 0.34\%& 0.54\%& 1.07\%& 1.70\%& 2.97\%& 12.53\%& 25.96\%& 37.25\%& \multicolumn{1}{|c}{10.3\%} \\
        \hline
        \multicolumn{1}{r|}{$I+A$} & 0.32\% (+4.5\%)& 0.51\% (+4.7\%)& 1.04\% (+2.3\%)& 1.62\% (+4.7\%)& 2.63\% (+11.4\%)& 12.38\% (+1.2\%)& 25.38\% (+2.2\%)& 36.18\% (+2.9\%)& \multicolumn{1}{|c}{10.0\% (+4.2\%)}\\
        \multicolumn{1}{r|}{$L+A$} & 0.34\% (-1.5\%)& 0.62\% (-15.9\%)& 1.15\% (-8.0\%)& 1.70\% (0.0\%)& 2.87\% (+3.4\%)& 12.01\% (+4.1\%)& 23.14\% (+10.9\%)& 33.64\% (+9.7\%)& \multicolumn{1}{|c}{9.4\% (+0.3\%)}\\
        \multicolumn{1}{r|}{$O+A$} & 0.32\% (+4.5\%)& 0.47\% (+12.1\%)& 0.81\% (+23.9\%)& 1.13\% (+33.5\%)& 2.23\% (+24.9\%)& 10.94\% (+12.7\%)& 23.78\% (+8.4\%)& 33.91\% (+9.0\%)& \multicolumn{1}{|c}{9.2\% (+16.1\%)}\\
        \multicolumn{1}{r|}{$R+A$} & 0.4\% (-19.4\%)& 0.67\% (-25.2\%)& 1.19\% (-11.7\%)& 1.91\% (-12.4\%)& 3.31\% (-11.4\%)& 13.58\% (-8.4\%)& 26.87\% (-3.5\%)& 37.81\% (-1.5\%)& \multicolumn{1}{|c}{10.7\% (-11.7\%)}\\
        \multicolumn{1}{r|}{$O_{oracle}+A$} & 0.09\% (+73.1\%)& 0.13\% (+75.7\%)& 0.15\% (+85.9\%)& 0.13\% (+92.4\%)& 0.36\% (+87.9\%)& 1.15\% (+90.8\%)& 2.29\% (+91.2\%)& 2.97\% (+92.0\%)& \multicolumn{1}{|c}{0.9\% (+86.1\%)}\\
        \multicolumn{1}{r|}{$O_{oracle,10}+A$} & 0.30\% (+10.4\%)& 0.23\% (+57.0\%)& 0.37\% (+65.3\%)& 0.37\% (+78.2\%)& 0.94\% (+68.4\%)& 5.19\% (+58.6\%)& 10.51\% (+59.5\%)& 15.38\% (+58.7\%)& \multicolumn{1}{|c}{4.2\% (+57.0\%)}\\
        \hline
        \multicolumn{1}{r|}{$I+L+A$} & 0.31\% (+7.5\%)& 0.58\% (-8.4\%)& 1.17\% (-9.9\%)& 1.71\% (-0.6\%)& 2.94\% (+1.0\%)& 11.67\% (+6.8\%)& 22.85\% (+12.0\%)& 33.22\% (+10.8\%)& \multicolumn{1}{|c}{9.3\% (+2.4\%)}\\
        \multicolumn{1}{r|}{$I+O+A$} & 0.21\% (+37.3\%)& 0.33\% (+38.3\%)& 0.68\% (+36.2\%)& 1.07\% (+37.1\%)& 2.28\% (+23.2\%)& 11.00\% (+12.2\%)& 23.22\% (+10.6\%)& 33.74\% (+9.4\%)& \multicolumn{1}{|c}{9.1\% (+25.5\%)}\\
        \multicolumn{1}{r|}{$O+L+A$} & 0.27\% (+19.4\%)& 0.43\% (+19.6\%)& 0.93\% (+12.7\%)& 1.09\% (+35.9\%)& 2.54\% (+14.5\%)& 11.40\% (+9.0\%)& 23.57\% (+9.2\%)& 33.61\% (+9.8\%)& \multicolumn{1}{|c}{9.2\% (+16.3\%)}\\
        \hline
        \multicolumn{1}{r|}{$I+O+L+A$} & 0.39\% (-16.4\%)& 0.59\% (-10.3\%)& 1.04\% (+2.3\%)& 1.42\% (+16.5\%)& 2.61\% (+12.1\%)& 10.85\% (+13.4\%)& 22.21\% (+14.4\%)& 32.18\% (+13.6\%)& \multicolumn{1}{|c}{8.9\% (+5.7\%)}\\
        \noalign{\hrule height 1.2pt}
    \end{tabular}
    }
    \vspace{-1.3em}
    \label{tab:3-eq-res-opt}
\end{table*}
%%%%%%%%%%%%%%% TABLE %%%%%%%%%%%%%%%%

\newcommand{\blocksubsubsection}[1]{\vspace{1em}\noindent\textbf{#1}\par\vspace{0.5em}}

%%%%%%%%%%%%%%%%% Figure %%%%%%%%%%%%%%%%%
%%%%%%%%%%%%%%%%% end figure  %%%%%%%%%%%%%

\subsection{Are more modalities always better for ASR?}
%  First, talk about average RB of +1, +2, and +3 modalities. 
% Then, talk about the consistency of +1, +2, and +3. 
%{\bf RQ: Are more modalities always better?}
Firstly, we examine the average performance, over all SNR levels, of adding more modalities. Compared to the audio-only baseline ($A\rightarrow T$) in the first row of Table \ref{tab:3-eq-res-opt}, the average RB values of adding one modality of either image (I), lip (L), or OCR (O) are all positive. Multimodal inputs also performed much better than any of the non-audio single input modalities.
When considering 3- or 4-modality combinations, both of them surpass the 2-modality models on average. The 4-modality model has the overall best average WER among all models, as shown in the last row in the table, while one 3-modality model has the highest RB.

Next, we  analyzed the consistency of adding modalities across all noise levels. When introducing one modality, adding lips does not bring a consistent benefit in noise, whereas adding images or OCR does. For 3-modality or 4-modality combinations, we observe more consistent benefits in noise. 
The 3-modality models are better than their corresponding 2-modality models at most noise levels. The 4-modality model outperforms the 3-modality models for SNR$<$0 but not for SNR from 0 to 20. This may be attributed to the longer input sequences, from which it is difficult for the model to find useful information. We discuss more about sequence noise caused by input length in Section \ref{subsec:length}. In general, we find evidence that multimodal ASR can benefit from at least 4 modalities as inputs, but the benefit varies with the noise level and thus how much complementary information is useful.

We found similar trends on SlideAVSR and S-MiT. On SlideAVSR, we use the OCR words and the first video frames as images because the videos in the dataset are about explaining slides. As shown in Table \ref{tab:slideavsr-res}, adding all OCR words ($O_{All}+A\rightarrow T$) and image ($I+A\rightarrow T$) has an average RB of +1.4\% and +5.3\%, respectively. On S-MiT, we extract video frames to be processed by visual encoders such as ViT and DALL-E. As shown in Table \ref{tab:smit-res}, the model achieves consistent improvements over the audio-only baseline when extra visual information is included. 
These demonstrate that the addition of supplementary visual modalities improves the overall recognition performance across datasets.

\begin{table}[t]
    \caption{WERs and relative benefits of adding modalities on SlideAVSR at different noise levels. The OCR words are ranked by FQ Ranker \cite{wang2024slideavsr}, and K indicates the maximum OCR word count.}
    \centering
    \setlength{\tabcolsep}{5pt}
    \resizebox{\columnwidth}{!}{%
    \begin{tabular}{rccccc}
        \noalign{\hrule height 1.2pt}
        \multicolumn{1}{r|}{\textbf{Inputs}} & +$\infty$ & 10 & 0 & -10& \multicolumn{1}{|c}{\textbf{Average}} \\
        \multicolumn{6}{c}{\cellcolor[HTML]{DEDCDC}WER $\downarrow$ (RB $\uparrow$)} \\
        \hline
        \multicolumn{1}{r|}{$A$} & 33.8\%& 42.5\%& 44.8\%& 70.6\%& \multicolumn{1}{|c}{47.9\%} \\
        \hline
        \multicolumn{1}{r|}{$O_{All}+A$} & 31.4\% (+7.0\%)& 37.2\% (+12.4\%)& 47.7\% (-6.4\%)& 75.9\% (-7.6\%)& \multicolumn{1}{|c}{48.0\% (+1.4\%)}\\
        \multicolumn{1}{r|}{$O_{K=30}+A$} & 30.5\% (+9.7\%)& 35.7\% (+15.8\%)& 46.9\% (-4.7\%)& 75.5\% (-6.9\%)& \multicolumn{1}{|c}{47.2\% (+3.5\%)}\\
        \multicolumn{1}{r|}{$O_{K=10}+A$} & 30.6\% (+9.5\%)& 34.6\% (+18.5\%)& 46.2\% (-3.1\%)& 77.9\% (-10.3\%)& \multicolumn{1}{|c}{47.3\% (+3.6\%)}\\
        \multicolumn{1}{r|}{$I+A$} & 28.5\% (+15.5\%)& 34.3\% (+19.3\%)& 46.1\% (-13.6\%)& 70.8\% (-0.2\%)& \multicolumn{1}{|c}{46.1\% (+5.3\%)}\\
        \noalign{\hrule height 1.2pt}
    \end{tabular}
    }
    \vspace{-1.0em}
    \label{tab:slideavsr-res}
\end{table}
%%%%%%%%%%%%%%% TABLE %%%%%%%%%%%%%%%%

%%%%%%%%%%%%%%% TABLE %%%%%%%%%%%%%%%%
\begin{table}[t]
    \caption{WERs and RB of adding modalities on S-MiT test set. $V$ refers to video modality, and $I$ refers to image.}
    \centering
    \setlength{\tabcolsep}{5pt}
    \resizebox{0.6\columnwidth}{!}{%
    \begin{tabular}{rccc}
        \noalign{\hrule height 1.2pt}
        \textbf{Inputs} & Vcodec & Visual Data & WER (RB $\uparrow$) \\
        \hline
        $A$ & - & - & 28.51\% \\
        \hline
        $V+A$ & ViT & 5 frames & 26.26\% (+7.9\%) \\
        $I+A$ & ViT & \engordnumber{1} frame & 26.71\% (+6.3\%) \\
        $I+A$ & DALL-E & \engordnumber{1} frame & 25.97\% (+8.9\%) \\
        \noalign{\hrule height 1.2pt}
    \end{tabular}
    }
    \vspace{-1.0em}
    \label{tab:smit-res}
\end{table}
%%%%%%%%%%%%%%% TABLE %%%%%%%%%%%%%%%%

\subsection{Acoustic noise: when is each modality most helpful?}
\label{subsec:acoustic_noise}
Modalities synchronized with audio (e.g., lip movement) could be effective in  different conditions than unsynchronized ones (e.g. image). 
As shown in Fig. \ref{fig:3eq_rb_combine}, for models with 2 modality inputs, the benefit of unsynchronized modalities follows a trend of increasing first and then decreasing, reaching a ``sweet spot'' in the middle. In contrast, the benefit of synchronized modalities grows stronger as the noise increases, similar to previous studies \cite{ma2023auto}. This discrepancy is because of how each visual modality aligns with the primary modality -- audio: when the audio is too noisy, the unsynchronized modalities (image, OCR) cannot establish a  correspondence with the audio, and  the visual information is useless.

When adding more than one modality, the model with additional image and lip information ($I+L+A$) displays a similar trend as  $L+A$ model; the $O+L+A$ model, on the other hand, is similar to $O+A$ model. This may indicate that the lip and OCR are more prominent modalities in the model decision. Meanwhile, the 4-modality model exhibits a trend similar to $O+A$ when there is less noise, and a trend similar to $L+A$ when there is more noise, having a sweet spot at SNR=2.5dB. This also supports the view that OCR and lip dominate the performance. 

\begin{figure}[t]
    \centering
    \includegraphics[width=0.6\columnwidth]{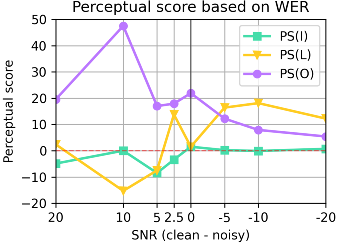}
    \vspace{-1.2em}
    \caption{Perceptual scores based on WER of each modality in models fine-tuned on 3-Equations. PS is computed on the 4-modality model ($I+O+L+A$).
    }
    \vspace{-1.2em}
    \label{fig:ps}
\end{figure}

We also compute each modality's Perceptual Score (PS) to better understand their contribution and effectiveness across noise levels. While we mostly find a similar PS trend to RB results, the fact that the PS(I) curve is near 0 for most SNR values suggests that, although adding the image information can improve ASR accuracy, the benefit might have more to do with regularization than complementary information. See Fig. \ref{fig:ps} and Supplementary Materials (Section IV) for more details.

%%%%%%%%%%%%%%%%% end figure  %%%%%%%%%%%%%

% To better understand the contribution and effectiveness of each modality across noise levels, we compute the Perceptual Score (PS) for each modality based on WER. 
% % Table \ref{tab:3eq-ps-shap} 
% Fig. \ref{fig:ps} presents the PS based on WER for individual modalities at varying noise levels. The PS is computed with Eq.(\ref{ps}) based on the 4-modality model ($I+O+L+A$) on 3-Equations, which reflects each modality's perceptual contribution. Overall, the OCR modality ($O$) consistently shows the highest positive PS across most noise levels, highlighting its dominance in model performance. Notably, $O$ achieves the highest PS at moderate noise levels, which aligns with our prior findings.
% The image modality ($I$) shows modest but generally stable contributions in PS, indicating that it plays a secondary but reliable role. The lip modality ($L$), while showing negative contributions in low-noise settings, exhibits an upward trend as noise increases, which is also consistent with our previous conclusions.

We observe some similar phenomena on SlideAVSR, as shown in Table \ref{tab:slideavsr-res}. For all three experiments that add OCR words and one experiment that add images, they all outperform the audio-only baseline at low noise levels (+$\infty$ and 10dB), but underperform the baseline at higher noise levels. They exhibit a sweet spot in the middle at 10dB, and this follows our conclusion that unsynchronized modalities need to establish a reliable correspondence with the audio to have an effect.

% \subsection{Proportion of Irrelevant Information (Sequence Noise)}
\subsection{Do irrelevant inputs (sequence noise) impact accuracy?}
\label{subsec:length}
%As mentioned above,  adding more modalities can potentially hurt the performance due to longer sequences, which are harder for the model to search for useful information.
We  investigated how the proportion of irrelevant information in the input sequence -- which we call \emph{sequence noise} -- can impact ASR accuracy. Recall that, in the 3-Equations dataset, each example contains 3 written equations, but only 2 of them are actually spoken. Therefore, the OCR modality inherently only contains around $\frac{2}{3}$ relevant information, and the model needs to learn -- by combining the OCR input with auditory information about what was actually said -- how to locate useful information from the input sequence. 

To create a scenario with higher sequence noise, we added to each example 7 extra irrelevant oracle OCR sentences that are randomly selected from other examples in addition to the original 3 sentences, resulting in a dataset with around $\frac{1}{5}$ relevant information. We then compared ASR accuracy of the two models (one with less, and one with more sequence noise).  Results are shown in Table \ref{tab:3-eq-res-opt} ($O_{oracle}+A$ and $O_{oracle, 10}+A$). We observe that including more sequence noise leads to much worse accuracy, with the overall average RB dropping from +86\% to + 57\%. In particular, the RB gap between the two models increases as the level of acoustic noise increases. 
These results may also explain why the 4-modality model is worse than the 3-modality or even 2-modality models under some conditions: adding less informative modalities introduces a longer sequence while reducing the proportion of relevant information in the sequence.

In SlideAVSR, we similarly explore the influence of sequence noise using OCR words. A sample slide in SlideAVSR can contain hundreds of OCR words, while only a small fraction of them are mentioned in the speech. According to our experiments on 3-Equations, it is conceivable that this information overload could harm performance. To evaluate this impact, we use FQ Ranker \cite{wang2024slideavsr} to filter the OCR words based on the frequency of word occurrences in Wikipedia, setting the maximum word count ($K$) to 10 and 30. As shown in Table \ref{tab:slideavsr-res} ($O_{All}$, $O_{K=30}$, $O_{K=10}$), although the performance of adding OCR is worse than the audio-only model at some noise levels, its overall RB increases as we filter more stringently.

%%%%%%%%%%%%%%% TABLE %%%%%%%%%%%%%%%%
\begin{table*}[t]
    \caption{WERs and relative benefits of adding modalities to the Mamba-based model on 3-Equations, at different noise levels.}
    \centering
    \setlength{\tabcolsep}{5pt}
    \resizebox{\textwidth}{!}{%
    \begin{tabular}{rccccccccc}
        \noalign{\hrule height 1.2pt}
        \multicolumn{1}{r|}{\textbf{Inputs}} & 20 & 10 & 5 & 2.5 & 0 & -5 & -10 & -20 & \multicolumn{1}{|c}{\textbf{Average}} \\
        \multicolumn{10}{c}{\cellcolor[HTML]{DEDCDC}WER $\downarrow$ (RB $\uparrow$)} \\
        \hline
        \multicolumn{1}{r|}{$A$} & 1.02\%& 1.37\%& 2.03\%& 2.50\%& 4.34\%& 15.13\%& 28.45\%& 38.18\%& \multicolumn{1}{|c}{11.6\%} \\
        \hline
        \multicolumn{1}{r|}{$I+A$} & 0.91\% (+10.8\%)& 1.34\% (+2.2\%)& 2.02\% (+0.5\%)& 2.76\% (-10.4\%)& 4.68\% (-7.8\%)& 15.54\% (-2.7\%)& 28.97\% (-1.8\%)& 39.10\% (-2.4\%)& \multicolumn{1}{|c}{11.9\% (-1.5\%)}\\
        \multicolumn{1}{r|}{$L+A$} & 1.31\% (-28.4\%)& 1.65\% (-20.4\%)& 2.48\% (-22.2\%)& 3.12\% (-24.8\%)& 4.94\% (-13.8\%)& 17.99\% (-18.9\%)& 31.88\% (-12.1\%)& 42.71\% (-11.9\%)& \multicolumn{1}{|c}{13.3\% (-19.1\%)}\\
        \multicolumn{1}{r|}{$O+A$} & 0.66\% (+35.3\%)& 0.82\% (+40.1\%)& 1.30\% (+36.0\%)& 1.80\% (+28.0\%)& 3.29\% (+24.2\%)& 12.31\% (+18.6\%)& 24.63\% (+13.4\%)& 33.44\% (+12.4\%)& \multicolumn{1}{|c}{9.8\% (+26.0\%)}\\
        \noalign{\hrule height 1.2pt}
    \end{tabular}
    }
    \vspace{-1.3em}
    \label{tab:3-eq-res-mb}
\end{table*}
%%%%%%%%%%%%%%% TABLE %%%%%%%%%%%%%%%%

\subsection{How do different visual representations impact accuracy?}

Next we explore the impact of the visual representation quality  on ASR accuracy. 
As shown in Fig. \ref{fig:datasets}, there are four image-based visual modalities in the 3-Equations dataset: image encoding, OCR text, oracle OCR text, and raster representation. These modalities can be placed along a spectrum from \textit{raw} to \textit{abstract} \cite{liang2024foundations}, where the image encoding is raw, and the OCR texts are more abstract as they present information extracted from images. As summarized in Table \ref{tab:3-eq-res-opt}, the average WER of 2-modality models combining visual modalities follows: $A$ (10.3)$>I+A$ (10.0)$>O+A$ (9.2)$>O_{oracle}+A$ (0.9). This can be attributed to the visual representations becoming more abstract and informative, thus reducing the visual noise level, making it easier for the model to utilize. Therefore, better visual representation with less visual noise can lead to better supplementary performance. We notice that adding raster has even worse accuracy than adding image, we suppose this is because: (1) the raster representation has a much longer token sequence (384 for image vs. 864 for raster) that hinders the model from finding useful information, as discussed in Section \ref{subsec:length}, and (2) the model still struggles in handling 2-D representations, even when it is as perfect as rasters. This suggests that more powerful positional encoding schemes than an absolute encoder might be necessary for 2-D representations.

Beyond the variations in visual representations, the capacity of the visual encoder itself may further influence the effective quality, or SNR, of the extracted visual features. We therefore investigate the impact of visual encoders by comparing models utilizing ViT and DALL-E on the S-MiT, the results are included in Table \ref{tab:smit-res}. From the results in the last two rows, DALL-E, with a relative benefit of +8.9\%, performs better than ViT on this task. This may be attributed to their distinct training objectives: ViT excels at extracting structural details, whereas DALL-E produces richer semantic features more aligned with language, thus facilitating multi-modal integration in our task.
We also note that $V+A$ outperforms $I+A$ when both use ViT as encoder, which indicates that DMLM is capable of learning from spatio-temporal information.

%%%%%%%%%%%%%%%%%%%%%%%%%%%%%%%%%%%%%%%%%%%%%%%%%%%%%%%%%%%%%%%%%%%%%%%%%%%%
% \section{Experiments: Model Type and Size}
\section{Experiments: Transformers vs. Mamba}

% \subsection{Selective State Space Models vs. Transformers}
Although Transformer-based MLLMs have achieved impressive performance, their quadratic-complexity attention is inefficient in terms of both latency and memory usage, especially for multimodal input streams which tend to be long. As an alternative method with fast inference speed and competitive accuracy, linear sequential state space models such as Mamba \cite{gumamba} are becoming more powerful and are growing in popularity for MLLMs in particular \cite{zhao2025cobra}.

%Meanwhile,  existing studies have been exploring extending Mamba LLMs into MLLMs to fully exploit its advantage. As widely observed, the computational complexity issue is aggravated in multi-modal tasks, because each modality introduces distinct representations to be jointly processed, leading to longer input sequences, larger model sizes, and heavier cross-modal attention computations. This can result in significantly higher memory usage and inference latency. Therefore, in addition to Transformer-based LLMs, we are also interested in examining the effectiveness and efficiency of multi-modal inputs for Mamba-based models.

To explore whether Mamba-based MLLMs exhibit some of the same trends regarding multimodality as Transformers do, we conduct similar 2-modality experiments ($I+A$, $L+A$, $O+A$) to compare a pretrained Mamba-130M to OPT-125M as the MLLM backbone. 
Results of the Mamba model on 3-Equations are summarized in Table \ref{tab:3-eq-res-mb}. By adding lip movements and OCR, we reach the same conclusion as OPT, that the modality synchronized with audio has increasing RB when there's more noise, while the unsynchronized modality has the greatest RB in moderate noise. At the same time, we also notice that all experiments that are conducted on Mamba perform worse than those on OPT. The RB of adding an image has a decreasing trend as the noise level goes up, which is different from the trend on OPT. 
We test the audio-only model and the $I+A$ model on clean audio (SNR=+$\infty$), where their WERs are 0.6\% and 0.94\%, respectively, resulting an RB of -10.6\% when adding $I$. 
This confirms that the general trend of adding images on Mamba is the same as OPT, but the sweet spot may be more architecture-dependent.

As a fundamental advantage of leveraging Mamba models, we want to examine the time cost of Mamba-based and Transformer-based multi-modal models. 
With our experimental configurations on 3-Equations, training Mamba for $A\rightarrow T$  takes about 197 seconds per epoch (sec/ep), while  an OPT model of similar size  takes  383 sec/ep.
With more input modalities, the length of the input sequence increases, which increases the time cost. In particular, for an $I+A\rightarrow T$ model, 
the training time of OPT grows to 305 sec/ep (55\% increase), while OPT it increased to 643 sec/ep (68\% increase). With three modalities ($I+L+A\rightarrow T$), the time cost further rises to 313 sec/ep (59\% total increase) and 837 sec/ep (219\% total increase), respectively. These indicate that Mamba-based models in general are more efficient, and are less affected by the increase in length. %This suggests that Mamba can be a potent alternative when the input length is longer with more modalities.
% We notice that the training time of OPT is degraded to 60\% when the second modality (image) is incorporated, while Mamba is decreased to 65\%. The speed of OPT and Mamba further drops to 46\% and 63\% when the third modality (lip) is introduced. These indicate that Mamba-based models in general are more efficient, and are less affected by the increase in length. This suggests that Mamba can be a potent alternative when the input length is long with more modalities.

On the other hand, our practical experiences with Mamba suggest it is not as stable as a Transformer-based model, since it seems to be sensitive to hyperparameters. For example, varying the learning rate for the $A\rightarrow T$ Mamba from 1e-3 to 1e-4 substantially increased the WER from 13.7\% to 20.4\%, but a comparable change in learning rate for OPT changed its accuracy only very slightly from 10.3\% to 10.6\%.  Furthermore, when adapting the model to accept multi-modal inputs, we find that different hyperparameters have to be used for different modality combinations to achieve reasonable performance. This may indicate that Mamba-based model may have high specialization but relatively low generalizability if the hyperparameters are optimal. 
We also experiment with different sizes of OPT and Mamba models, and observe that they exhibit different patterns when scaling up the model size. See details in Supplementary Materials (Section V).

% \subsection{Model Size}
% In order to compare the performance impact of different model size, we conduct the A$\rightarrow$T and I+A$\rightarrow$T experiments on OPT-350M and Mamba-370M using the same configurations. For a fair comparison, we compare the larger models to off-the-shelf OPT-125M and Mamba-130M model (without our pre-training stage). From the results, we find that for OPT models, a larger model doesn't consistently  outperform the smaller baseline, whereas for Mamba model the larger model is generally better. We suspect this is because the larger OPT models has better pre-trained text modeling, which might help the model to predict text tokens in the inference phase, especially for high noise scenarios. While for the Mamba model, it shows rational scaling advantage with more parameters.

%%%%%%%%%%%%%%%%%%%%%%%%%%%%%%%%%%%%%%%%%%%%%%%%%%%%%%%%%%%%%%%%%%%%%%%%%%%%
\section{Experiments: Input Format and  Loss Weights}
With MLLMs,  practical questions arise of (1) how to format the input stream so that the model can find what it needs, and (2) how much to weigh each modality in the  loss function. %We explore these questions below.

\vspace{-0.7em}
\subsection{Multi-modal inputs: interleaved or blocked?}
Interleaved inputs are common in vision-language tasks, as this format is ubiquitous in web-crawled data. 
Previous works have examined the benefit of using interleaved data, and find it endows MLLMs with in-context few-shot learning capabilities \cite{tian2024mm, alayrac2022flamingo}. 
However, there has been little exploration of interleaving modalities when both modalities contain dense and synchronized information.  
We thus compared interleaved to non-interleaved (blocked) input formats with $V+A$ task. Specifically, 
%the impact of interleaved speech-video (image) inputs. The basic idea is to use the interleaved sequence of video frames and audio clips as an input sequence. We suspect this may help the model more easily correspond to the speech and the video part. 
we conducted experiments on the S-MiT by interleaving the 5 video frames evenly with audio. I.e., if the speech contains 300 audio tokens, the final input sequence will be like: ``\textless Frames [0]\textgreater \textless AudioTokens [0:60]\textgreater \textless Frames [1]\textgreater \textless AudioTokens [60:120]\textgreater \ldots \textless Frames [4]\textgreater \textless AudioTokens [240:300]\textgreater". Using ViT as visual encoder to tokenize video frames, the model achieves a WER of 28.4\% in the interleaving manner, which is +0.4\% RB to the audio-only model. Although this model is better than the audio-only model, its performance is still much worse than the non-interleaved (blocked) input.

\subsection{In what order should the different modalities be input?}
%%%%%%%%%%%%%%%%% Figure %%%%%%%%%%%%%%%%%
\begin{figure}[t]
    \centering
    \includegraphics[width=0.6\columnwidth]{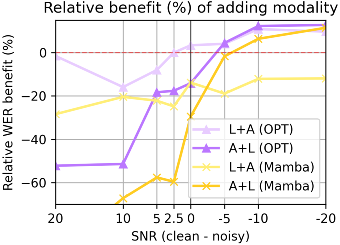}
    \vspace{-1em}
    \caption{RB of audio-first or audio-later inputs on OPT and Mamba model.}
    \vspace{-1.0em}
    \label{fig:input_order}
\end{figure}
%%%%%%%%%%%%%%%%% end figure  %%%%%%%%%%%%%

Previous work studying multi-modal inputs observed order sensitivity in MLLMs, and found that prompt orders and modality orders can significantly affect results. Some demonstrate that MLLMs have a preference for the beginning and end of contexts, and placing important content in these positions can enhance performance \cite{tan2024order}. Others examine the impact of modality sequencing in prompt context, and found the impact is context-dependent and  related to task complexity \cite{wardle2025image}.  

We explored the impact of switching modality input orders in 2-modality models using either Transformer or Mamba as backbone. 
Specifically, we compared the accuracy of a model $L+A\rightarrow T$ to $A+L\rightarrow T$, i.e., the order of the two inputs (audio, lips) are swapped. Results are shown in Figure \ref{fig:input_order}. They indicate that RB of the audio-first models keeps increasing as the noise level increases (lower SNR) and eventually outperforms audio-later models, and that audio-first is more beneficial in noisy situations. We found similar trends for other 2-modality models.

We speculate that the results are due to an interaction between three factors: (1) the absolute positional encoder that spans the entire input sequence; (2) primacy/recency effects of how Transformers sometimes prefer to attend to tokens near the start/end (respectively) of the input  \cite{tan2024order}; and (3) how the audio noise in 3-Equations  was added to the \emph{second} half  of each utterance. 
With the $A+L \rightarrow T$ model, the clean part (first half) of the first utterance is at a known location (i.e., the start of the input sequence) so that the model can easily find it via the absolute positional encoding. In contrast, with the $L+A \rightarrow T$ model, the audio begins only after a \emph{variable-length} lip sequence, and hence the clean part of the first utterance is difficult for the model to find. Moreover, the most recent part of the audio is the noisy half and thus, especially at low SNR, not very useful. See Supplementary Materials (Section VI) for more explanation.

%, but they do not offer a solid explanation of why this phenomenon happens. 
%We hypothesize that the accuracy will be improved if the audio is clean and can easily be found at a fixed location (e.g., the beginning) or occurs very recently.
%Hence, due to our method of adding noise (2-noise), inputting audio first can help the model attend more to clean audio at the beginning, especially in noisy conditions. However, RB of $A+I$ on OPT exhibits a different trend that first decreases below zero, then starts to increase above zero. Image modality shows a different trend since it has a fixed length while the other modalities do not, which may help the model find the beginning of the audio even if it appears later. We suspect this trend has to do with the image providing the greatest benefit in moderate noise.

\subsection{How to weigh the different multimodal loss components?}
\label{subsec:modal_loss}
%%%%%%%%%%%%%%% TABLE %%%%%%%%%%%%%%%%
\begin{table}[t]
    \caption{Evaluate WER(\%) on the 3-Equations of all noise levels (SNRs) with different modality loss weight configurations. Loss weight is displayed in the order of (speech, text, image). Only text weight is non-zero in the second experiment.}
    \centering
    \renewcommand{\arraystretch}{1.2}
    \setlength{\tabcolsep}{5pt}
    \resizebox{0.9\columnwidth}{!}{%
    \begin{tabular}{r|c|cccccccc}
        \noalign{\hrule height 1.2pt}
        \textbf{Task} & \textbf{Loss} & 20 & 10 & 5 & 2.5 & 0 & -5 & -10 & -20 \\
        \hline
        $A\rightarrow T$ & (0.3,0.5,0.5) & 0.3\%& 0.5\%& 1.1\%& 1.7\%& 3.0\%& 12.5\%& 26.0\%& 37.0\%\\
        $A\rightarrow T$ & (0.0,0.5,0.0)&  0.3\%&  0.6\%&  1.2\%&  1.8\%&  3.3\%&  13.5\%&  27.0\%&  38.4\%\\
        \noalign{\hrule height 1.2pt}
    \end{tabular}
    }
    \vspace{-1.2em}
    \label{tab:loss_weight_text}
\end{table}
%%%%%%%%%%%%%%% TABLE %%%%%%%%%%%%%%%%

During the training stage of a Transformer-based model, the loss function is used to predict the next token as the autoregressive objective. Hence, a modality-weighted loss can be viewed as how well we want the model to learn to generate tokens of that modality. Since in the ASR tasks we only care about the transcription, which is merely text generation, we wonder if setting the loss weights of other modalities to zero can lead to good results with faster convergence. However, experimental results in Table \ref{tab:loss_weight_text} show that this is not the case. This may be because the model needs to learn the internal logic or structure of that modality, and such tasks can be viewed as an intrinsic multitask learning setting (e.g., the speech-to-speech next token prediction task contributes to the speech-to-text task). This suggests that, even for simple speech-to-text generation tasks on a pretrained MLLM, learning from modalities other than text remains important.

\vspace{-0.7em}
\section{Conclusion and Future Research}

This work provides an in-depth investigation into how multiple input modalities impact  speech recognition accuracy in MLLMs. Our experiments suggest that: 
(1) Fusing more modalities usually enhances performance, but the benefit of each modality depends on the auditory noise. Specifically, synchronized modalities become more influential at high noise levels, whereas unsynchronized modalities provide the greatest benefit at moderate (``sweet spot'') noise levels. 
(2) Adding irrelevant information into the input sequence decreases accuracy. 
(3) Visual representations with higher-quality and generated from better visual encoders consistently improve accuracy. 
Also, the performance gap between raster and oracle OCR (both are essentially perfect representations) suggests that  MLLMs with absolute positional encoders struggle to handle 2-D representations. 
(4) Mamba shows similar trends as Transformers with faster speed, but are less stable to train. 
(5) Different input formats may vary in effectiveness, suggesting new directions for optimizing multi-modal input formatting.

%One limitation of our RB metric is that when considering the regularization effect, the PS can be more effective, because PS can eliminate the regularization effect with minimal domain shift.
%Another limitation is that using a synthesized dataset may introduce bias due to inherent synthesis errors of the speech and lip movements. 
%We also have reservations about the potential enhancement of inference speed with multi-modal inputs. 
% In future research, more modalities, model architectures, and real-world datasets should be explored. 

Our results suggest some potential research directions to improve multimodal ASR: (1)  Develop more powerful  positional encodings that help the model to collate information across modalities and also to handle 2-D visual representations; (2) Develop stronger visual encoders that can extract more detailed (e.g., text content) from images; (3) Explore more recent state space models that might obtain the ``best of both worlds'' of fast inference and high accuracy. 

{\bf Acknowledgement}: This research was supported by the NSF National AI Institute for Student-AI Teaming (iSAT) under grant DRL \#2019805, and also from an NSF CAREER grant \#2046505. The opinions expressed are those of the authors and do not represent views of the NSF.

%\section*{Acknowledgments}

%%%%%%%%%%%%%%%%% APPENDIX %%%%%%%%%%%%%%%%%

% \appendix[Proof of the Equations]
% Proof

%%%%%%%%%%%%%%%%% REFERENCES %%%%%%%%%%%%%%%%%

% \small

\bibliographystyle{IEEEtran}
\bibliography{ref}

\newpage

%%%%%%%%%%%%%%%%% BIOGRAPHY %%%%%%%%%%%%%%%%%

 

% \end{IEEEbiography}



% \vfill

\end{document}